\documentclass[
aps,
prl,
showpacs
amsmath,
amssymb,
floatfix,
longbibliography,
letterpaper,
lengthcheck,
superscriptaddress
]{revtex4-1}

\usepackage{graphicx}
\usepackage{rotating}
\usepackage{bbm}
\usepackage{subfigure}
\usepackage{color}
\usepackage{hyperref}
\usepackage[all,warning]{onlyamsmath}
\usepackage{amsfonts}
\usepackage{braket}

\graphicspath{
  {figures/}
}

\def\imagi{\mathrm{i}}

\def\a{\hat{a}}
\def\ad{\hat{a}^{\dagger}}

\begin{document}

\title{Kohn-Sham Approach to Quantum Electrodynamical Density Functional Theory: Exact Time-Dependent Effective Potentials in Real Space}

\author{Johannes Flick}
\email[]{flick@fhi-berlin.mpg.de}
\affiliation{Fritz-Haber-Institut der Max-Planck-Gesellschaft, Faradayweg 4-6, D-14195 Berlin-Dahlem, Germany}
\author{Michael Ruggenthaler}
\email[]{michael.ruggenthaler@uibk.ac.at}
\affiliation{Institut f\"ur Theoretische Physik, Universit\"at Innsbruck, Technikerstra{\ss}e 21A, A-6020 Innsbruck, Austria}
\affiliation{Max Planck Institute for the Structure and Dynamics of Matter and Center for Free-Electron Laser Science \& Department of Physics, Luruper Chaussee 149, 22761 Hamburg, Germany}
\author{Heiko Appel}
\email[]{appel@fhi-berlin.mpg.de}
\affiliation{Fritz-Haber-Institut der Max-Planck-Gesellschaft, Faradayweg 4-6, D-14195 Berlin-Dahlem, Germany}
\affiliation{Max Planck Institute for the Structure and Dynamics of Matter and Center for Free-Electron Laser Science \& Department of Physics, Luruper Chaussee 149, 22761 Hamburg, Germany}
\author{Angel Rubio}
\email[]{angel.rubio@mpsd.mpg.de}
\affiliation{Fritz-Haber-Institut der Max-Planck-Gesellschaft, Faradayweg 4-6, D-14195 Berlin-Dahlem, Germany}
\affiliation{Max Planck Institute for the Structure and Dynamics of Matter and Center for Free-Electron Laser Science \& Department of Physics, Luruper Chaussee 149, 22761 Hamburg, Germany}
\affiliation{Nano-Bio Spectroscopy Group and ETSF,  Dpto. Fisica de Materiales, Universidad del Pa\'is Vasco, 20018 San Sebasti\'an, Spain}

\date{\today}

\begin{abstract}
The density-functional approach to quantum electrodynamics is extending traditional density-functional theory and
opens the possibility to describe electron-photon interactions in terms of effective Kohn-Sham potentials. In this work,
we numerically construct the exact electron-photon Kohn-Sham potentials for a prototype system which 
consists of a trapped electron coupled to a quantized electromagnetic mode in an optical high-Q cavity. 
While the effective current that acts on the photons is known explicitly, the exact effective potential 
that describes the forces exerted by the photons on the electrons is obtained from a fixed-point inversion 
scheme. This procedure allows us to uncover important beyond-mean-field features of the effective potential which 
mark the breakdown of classical light-matter interactions. We observe peak and step structures in the effective potentials, 
which can be attributed solely to the quantum nature of light, i.e., they are real-space signatures of the photons. 
Our findings show how the ubiquitous dipole interaction with a classical electromagnetic field has to
be modified in real-space in order to take the quantum nature of the electromagnetic field fully into account.\\
\end{abstract}

%\pacs{31.15.ee}{Time-dependent density functional theory}
%\pacs{31.10.+z}{Theory of electronic structure, electronic transitions, and chemical binding}
%\pacs{71.15.Mb}{Density functional theory, local density approximation, gradient and other corrections}

\maketitle

\begin{figure}[t]
  \begin{center}  
    \includegraphics[width=0.4\textwidth]{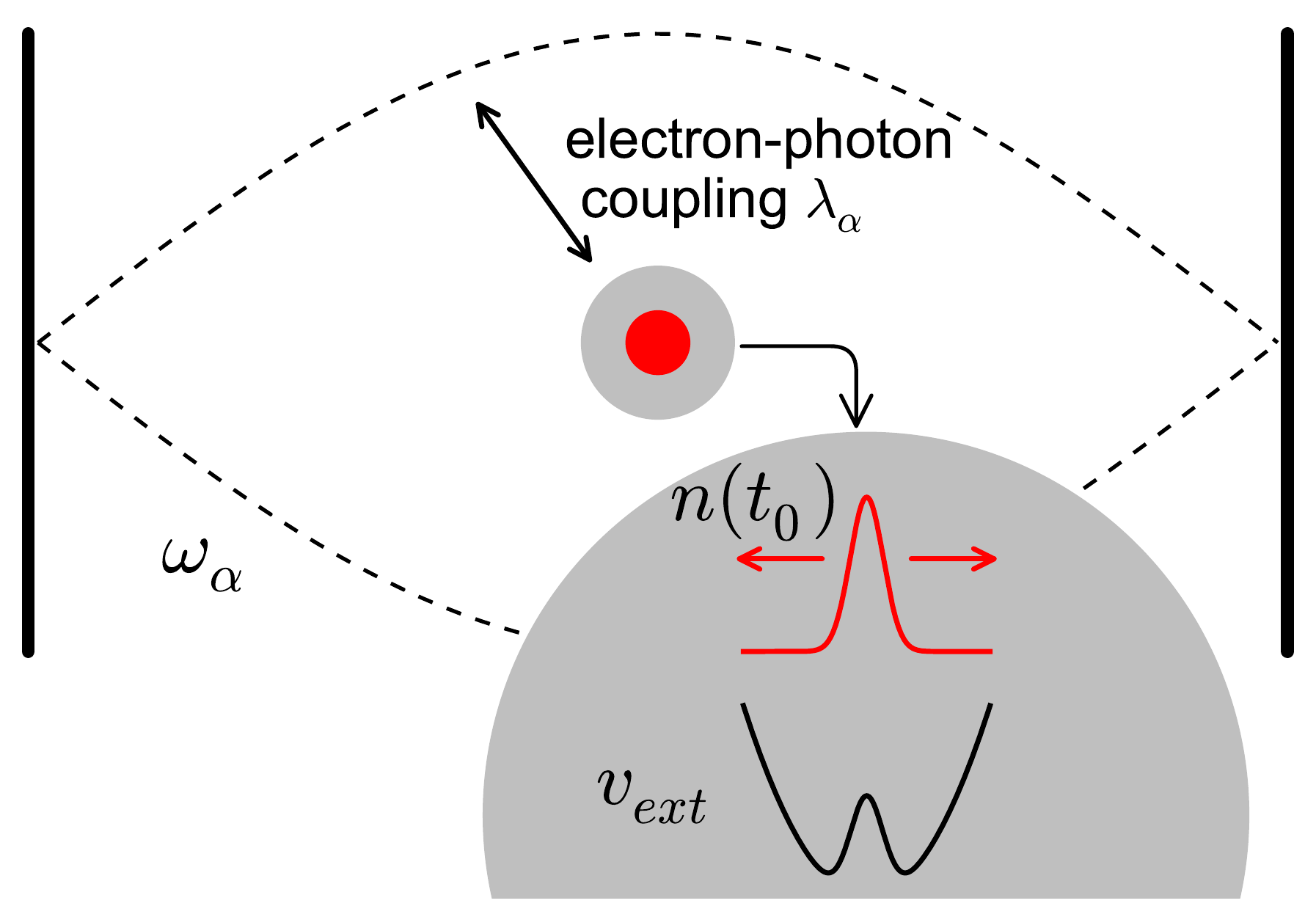}
  \end{center}
  \caption{{The figure schematically illustrates a} two-dimensional {optical} cavity containing one atom, with a single electron. The {coupling of the electron to the} cavity {mode at} resonance frequency $\omega_\alpha$ and {with} electron$-$photon coupling strength $\lambda_\alpha$ modifies the dynamics of the electron density $n(\textbf{r}t)$, which moves in the external potential $v_\text{ext}(\textbf{r}t)$.}
  \label{fig:model} 
\end{figure}

In the last decades, the quantum nature of light has inspired many experimental and theoretical developments in physics. In particular, the fields of cavity~\cite{kippenberg-2008} and circuit~\cite{schoelkopf-2008} quantum electrodynamics (QED) have recently seen exceptional progress. For instance, slow photons in vacuum~\cite{giovannini-2015} and two-ion superradiant states~\cite{casabone-2015} have been observed, and only recently the chemical landscape of a molecule has been modified using strong coupling to photons \cite{schwartz-2011, hutchison-2012}, which can be also termed QED chemistry.\\
However, traditional ab-initio approaches developed to investigate large quantum systems (see, e.g.,~\cite{fetter-2003,  marques-2012, ullrich-2012, stefanucci-2013, Bartlett-2007,Thomas-2015}) are not fully applicable in situations where the quantum nature of light becomes important. These many-body methods ignore {typically} the quantum-mechanical coupling to photons and {usually} take {only} the classical Coulomb interaction into account. Recently, an approach that treats particles (electrons, ions) and the photons on equal footing and closes the gap between traditional many-body and quantum-optical methods has been proposed~\cite{ruggenthaler-2011b,tokatly-2013,ruggenthaler-2014}. This so-called quantum-electrodynamical density-functional theory (QEDFT) allows to represent the coupled particle-photon system by two uncoupled, yet non-linear {auxiliary quantum} systems. The resulting multi-component Kohn-Sham systems are subject to effective potentials that take into account the particle-particle (Coulomb) interaction and the particle-photon interaction. If we employ approximations to this new type of Kohn-Sham potentials, the resulting equations become numerically feasible and ab-initio calculations of large quantum systems coupled to photons are possible \cite{marques-2003, castro-2006, andrade-2014}. While a wealth of approximations to the particle-particle interaction part of the effective potential are known (see, e.g.,~\cite{gross-1995, marques-2012, ullrich-2012}) at the moment there is only one approximation for the particle-photon part of the effective potential beyond the classical mean-field approximation available~\cite{pellegrini-2015}. Indeed, besides its existence {and uniqueness}, so far nothing is known about the exact real-space properties of this particle-photon effective potential and how it models the interaction between charged particles and {quantized photon fields}.  
\\ 
In this letter, we present the exact space- and time-resolved Kohn-Sham potential for a coupled multi-component electron-photon system in an optical cavity. The prototype system {that we consider in the present work} is a two-dimensional quantum ring containing one electron that is coupled to a single photon mode. We construct the exact effective potential by a fixed-point procedure~\cite{ruggenthaler-2011a,ruggenthaler-2012, nielsen-2013, nielsen-2015} and study ground-state properties as well as the time-evolution of the electron-photon system. In the first case we identify electronic states in the weak and strong-coupling regime, which cannot be generated by any classical light field in dipole approximation. In the time-dependent framework, we analyze the quantum effects of the photon field by putting the field initially into (1) a coherent state and (2) a linear superposition of two Fock number states and show when the {quantum nature} of the photon field {induces} the dominant contribution to the Kohn-Sham electron-photon exchange-correlation (xc) potential. Here we find that the electron-photon interaction {is responsible for} steps and peaks {in} the {exact} Kohn-Sham potential. {Similar steps and peaks have been found} in purely electronic time-dependent density functional theory (TDDFT) {in} e.g., charge-transfer processes~\cite{elliott-2012,fuks-2013}, but these exact features have so far only been observed for the time-dependent case in one-dimensional models \cite{elliott-2012}.\\
The static and dynamical behavior of the coupled electron-photon systems {that we consider in the present work} {is given by the following Hamiltonian}~\cite{tokatly-2013,ruggenthaler-2014}, 
\begin{align}
\label{eqn:H} 
\hat{H}&=\sum_i\left(-\frac{\hbar^2}{2m}\vec{\nabla}_i^2+v_\text{ext}(\textbf{r}_it)\right)+\frac{e^2}{4 \pi\epsilon_0}\sum_{ij,i>j}\frac{1}{\left|\textbf{r}_i-\textbf{r}_j\right|}\nonumber\\
&+\sum_{\alpha}\frac{1}{2}\left[\hat{p}^2_{\alpha}+\omega^2_{\alpha}\left(\hat{q}_{\alpha}-\frac{\boldsymbol{\lambda}_{\alpha}}{\omega_{\alpha}} \cdot \textbf{R} \right)^2\right]+\frac{j^\alpha_\text{ext}(t)}{\omega_\alpha}\hat{q}_\alpha.
\end{align}
The first part of the Hamiltonian describes the electronic subsystem and contains the non-relativistic kinetic energy, the external-potential energy due to a classical external potential $v_{\text{ext}}(\textbf{r}t)$ and the Coulomb-interaction energy. The second part of the Hamiltonian accounts for the presence of the photons, where the electron-photon interaction is described in dipole approximation, i.e, the dipole-moment operator $\textbf{R} = \sum_i \textbf{r}_i$ couples linearly to the photon displacement coordinate $\hat{q}_\alpha = \sqrt{\frac{\hbar}{2\omega_\alpha}}\left(\ad_\alpha + \a_\alpha \right)$, which is proportional to the {quantized displacement} field 
component of the $\alpha$th mode, {i.e, $\hat{\textbf{D}}_\alpha = {\epsilon_0\omega_\alpha}\boldsymbol{\lambda}_{\alpha}\hat{q}_\alpha/e $.} {Although we restrict ourselves here to dipole-coupling,} all findings shown below also apply for beyond-dipole situations. The electron-photon coupling is given by ${\boldsymbol \lambda_\alpha} = \lambda_{\alpha} \mathbf{e}_{\alpha}$, where $\mathbf{e}_{\alpha}$ is the polarization vector of the photon field. 
(For later reference, we measure $\lambda_\alpha$ in units of $[{\sqrt{\text{meV}}}/{\text{nm}}]$). In addition, the Hamiltonian contains a quadratic electron self-interaction $\boldsymbol{\lambda}^2_{\alpha}\textbf{R}^2/\omega^2_{\alpha} $ and the photons interact furthermore with a classical external current $j^{\alpha}_{\text{ext}}(t)$. {We emphasize that} both external potentials, i.e., $v_{\text{ext}}(\textbf{r}t)$ and $j^{\alpha}_{\text{ext}}(t)$, can be used to control the quantum system.\\
%\Radd{(In SI units, this does not have the dimension of a current. Indeed we have $[kg^{1/2} m s^{-3}]$. Should we remark on that?)}\\
In QEDFT the electron-photon system is exactly described by two reduced quantities that couple to the external control fields~\cite{ruggenthaler-2011b,tokatly-2013,ruggenthaler-2014}. In the case of the Hamiltonian (\ref{eqn:H}) these reduced quantities are the {usual electronic} density $n(\textbf{r}t) = \langle \hat{n}(\textbf{r})\rangle$, where $\hat{n}(\textbf{r})=\sum_i\delta(\textbf{r}-\textbf{r}_i)$, and the expectation value of the photon coordinates $q_\alpha(t) = \langle \hat{q}_{\alpha} \rangle$. In principle, we only need to calculate these expectation values and can then determine (for a fixed initial state) all properties of the electron-photon system. To calculate $n(\textbf{r}t)$ and $q_\alpha(t)$ one only needs to solve the corresponding coupled equations of motions for these two basic variables in the system, i.e., the Ehrenfest equations~\cite{tokatly-2013,ruggenthaler-2014}
\begin{align} 
\label{eq:eom-photon}
\partial^2_t{q}_\alpha(t) & = \omega_\alpha {\boldsymbol \lambda_\alpha} \cdot \textbf{R}(t)- \omega^2_\alpha  q_\alpha(t) - \frac{j^\alpha_{\text{ext}}(t)}{\omega_\alpha}, \\
\label{eq:eom-electron}
\partial^2_t  n(\textbf{r}t) &= \vec{\nabla} \cdot \textbf{Q}(\textbf{r}t) + \frac{1}{m}  \vec{\nabla}\cdot\left(n(\textbf{r}t)\vec{\nabla} v_{\text{ext}}(\textbf{r}t)\right)\nonumber\\
&+ \sum_\alpha {\boldsymbol \lambda_\alpha} \cdot \vec{\nabla} \bra{\Psi} \hat{n}(\textbf{r}) \left(   {\boldsymbol\lambda_\alpha} \cdot \textbf{r}  - \omega_\alpha \hat{q}_\alpha\right)\ket{\Psi},\nonumber\\ 
\end{align}
with the local-force density of the electrons given by ${Q}_k(\textbf{r}t) = \sum_l\partial_l T_{kl}(\textbf{r}t) + W_k(\textbf{r}t)$, where the first term describes the momentum-stress forces and the second term is responsible for the forces due to the particle-particle interactions. In order to solve these implicit equations, we would need to find explicit expressions in terms of $n(\textbf{r}t)$ and $q_\alpha(t)$ for the different force densities. We note, that in the equations of motion for the photon coordinates all terms are explicitly known, and hence the unknown expressions that take care of the proper description of the electron-photon interactions are contained solely in the electronic equation. To make approximations for these unknown quantities easier, one can adopt a Kohn-Sham scheme, such that approximations in terms of the force densities of the uncoupled and non-interacting system become possible. This approach has been applied highly successfully to electronic-structure calculations (see, e.g.,~\cite{gross-1995, marques-2012, ullrich-2012}). In a Kohn-Sham approach to the electron-photon system the missing forces are accounted for by the effective potential {$v_s(\textbf{r}t)$} that naturally splits into two parts $v_s(\textbf{r}t)= v_{\text{ext}}(\textbf{r}t) + v_{\text{Mxc}}(\textbf{r}t)$, where $v_{\text{ext}}$ is the external potential of the original problem, and $v_{\text{Mxc}}$ (Mean-field-exchange-correlation) denotes the effective potential due to the interaction with the photons. {In general, $v_s(\textbf{r}t)$ contains both, {the} contributions due to the Coulombic electron-electron repulsion in the original many-body problem, and {in addition the contributions} from the electron-photon interaction, here in dipole approximation. However,} since our aim is to investigate the effective potentials due to the coupling between photons and electrons, we restrict ourselves {here} to a single electron in an semiconductor GaAs quantum ring, which is placed in a cavity and is assumed to couple to a specific cavity photon mode, as depicted schematically in Fig.~\ref{fig:model}.
This restriction to a single electron has the advantage that we only have contributions of the electron-photon interaction in the effective Kohn-Sham potential $v_s(\textbf{r}t)$ and
can exclusively study only their behavior. We treat the electron in this semiconductor medium by using an effective mass $m=0.067m_e$ \cite{rasanen-2007} and we employ ${v}_\text{ext}(\textbf{r}) = \frac{1}{2}m\omega_0^2 \textbf{r}^2 + V_0 e^{-\textbf{r}^2/d^2}$, where $\textbf{r}^2=x^2+y^2$, to describe the quantum ring. This potential effectively confines the electron in a harmonic {trap}, which contains a Gaussian peak in the center. For the system at hand, we choose the experimental parameter values \cite{rasanen-2007} $\hbar\omega_0 = 10\text{meV}$, $V_0 = 200\text{meV}$, $d=10\text{nm}$, and the effective {di}electric constant $\kappa=12.7\epsilon_0$. The two-dimensional electronic system has a non-degenerate ground state and a two-fold degeneracy in the excited states \cite{rasanen-2007}. We choose the photon frequency $\omega_\alpha$ in resonance with the transition between the ground state and the first excited state in the electronic system. Thus, ${\hbar}\omega_\alpha = 1.41\text{meV}$. For simplicity, we restrict {ourselves} to one of the two independent polarization directions of the field mode and use $\mathbf{e}_{\alpha} = (1,1)$ without loss of generality.
\\
The Kohn-Sham scheme then decouples the two subsystems, which leads us to two evolution equations of the form  
\begin{align} 
\label{eq:ks-electron}
\imagi \hbar \partial_t \phi(\textbf{r}t) = -\frac{\hbar^2}{2m}\vec{\nabla}^2 \phi(\textbf{r}t) + v_{s}(\textbf{r}t)\phi(\textbf{r}t),
\end{align}
\begin{align} 
\label{eq:ks-photon}
\imagi \hbar \partial_t \ket{\alpha,t} = \frac{1}{2}\left[\hat{p}_\alpha^2 + \omega_\alpha^2 \hat{q}_\alpha^2\right]\ket{\alpha,t} +  \frac{j_\text{s}^\alpha(t)}{\omega_\alpha}\hat{q}_\alpha \ket{\alpha,t},
\end{align}
where the Kohn-Sham photon wavefunction is given by $\ket{\alpha,t} = \sum_n c_n \ket{n,t}$ and $\ket{n,t}$ are the {Fock} number states of cavity mode $\alpha$. The Kohn-Sham construction furthermore requires that the initial state of the Kohn-Sham system has to have the same density $n(\textbf{r},0)$ and time-derivative $\dot{n}(\textbf{r},0)$ as the coupled system. The same is required for the basic variable in the photon system, i.e.,  $q_\alpha(0)$ and $\dot{q}_\alpha(0)$.
\\
To determine the in general unknown effective potential $v_s(\textbf{r}t)$ in terms of $n(\textbf{r}t)$ and $q_\alpha(t)$, we use a fixed-point method originally developed for purely electronic TDDFT~\cite{ruggenthaler-2011a,ruggenthaler-2012, nielsen-2013, nielsen-2015}. Although naively one could expect that a fixed-point iteration is also needed to determine the effective current $j_\text{s}^\alpha(t)$, from Eq.~(\ref{eq:eom-photon}) this current is known explicitly, i.e., $j_\text{s}^\alpha(t) = j_\text{ext}^\alpha(t) + \omega^2_\alpha {\boldsymbol{\lambda}}_\alpha \cdot \textbf{R}(t)$. Hence we only need to determine $v_s(\textbf{r}t)$, for which we use the fixed-point formula 
\begin{align}
\label{eq:iteration}
-\vec{\nabla} \cdot \left( n(\textbf{r}t) \vec{\nabla} v_{k+1}(\textbf{r}t) \right) &= \partial_t^2 \left[ n([v_k],\textbf{r}t)-n(\textbf{r}t) \right]\nonumber\\
&-\vec{\nabla} \cdot \left(n([v_k],\textbf{r}t)\vec{\nabla}v_k(\textbf{r}t) \right).
\end{align}
To find the fixed-points of this equation, we solve first the Schr\"odinger equation for $v_k$ (using zero-boundary conditions) and from the exact many-body solution we determine the corresponding $n[v_k]$. Next, we employ a multigrid solver to invert the Sturm-Liouville problem in Eq.~\ref{eq:iteration}, which yields $v_{k+1}$. This procedure is repeated until convergence to the fixed-point has been reached. To speed up convergence, we employ a direct inversion in the iterative subspace (DIIS) approach~\cite{garza-2012}. We have tested the validity of this approach in the time-independent situation  by comparing to the well-known analytic inversion formula for one-electron and two-electron singlet problems~\cite{helbig-2009} 
\begin{align}  
v_{s}^{(0)}(\textbf{r}) = \frac{\hbar^2}{2m}\frac{\vec{\nabla}^2\sqrt{n_0(\textbf{r})}}{\sqrt{n_0(\textbf{r})}} + E_0.
\end{align}
First, let us investigate the effective potential in the case of the ground state of the multi-component system. Using exact diagonalization~\cite{flick-2014}, we are able to calculate the exact ground state of the correlated electron-photon system. We use a 127x127 two-dimensional real-space grid for the electron and 40 photon number states. This amounts to a dimensionality of the full problem of 127x127x40 = 645160 basis functions. From this we then determine with the above iteration procedure the exact $v_\text{s}$. Since we treat the interaction with the photon field in dipole approximation, the classical field contribution to the effective potential is $v_\text{M}(\textbf{r}t) = \omega_\alpha {\boldsymbol \lambda}_{\alpha} q_\alpha(t) \cdot \textbf{r}$. If the impact of the quantum nature of the cavity light field would be negligible, the classical field would be the only contribution. Thus, if we use that $v_\text{Mxc} = v_\text{M} + v_\text{xc}$, the non-dipole corrections (to all orders in $\mathbf{r}$) due to $v_\text{xc}$ are a direct measure of the non-classical light-matter interaction.
\begin{figure}[ht]
  \begin{center}  
    \includegraphics[width=0.49\textwidth]{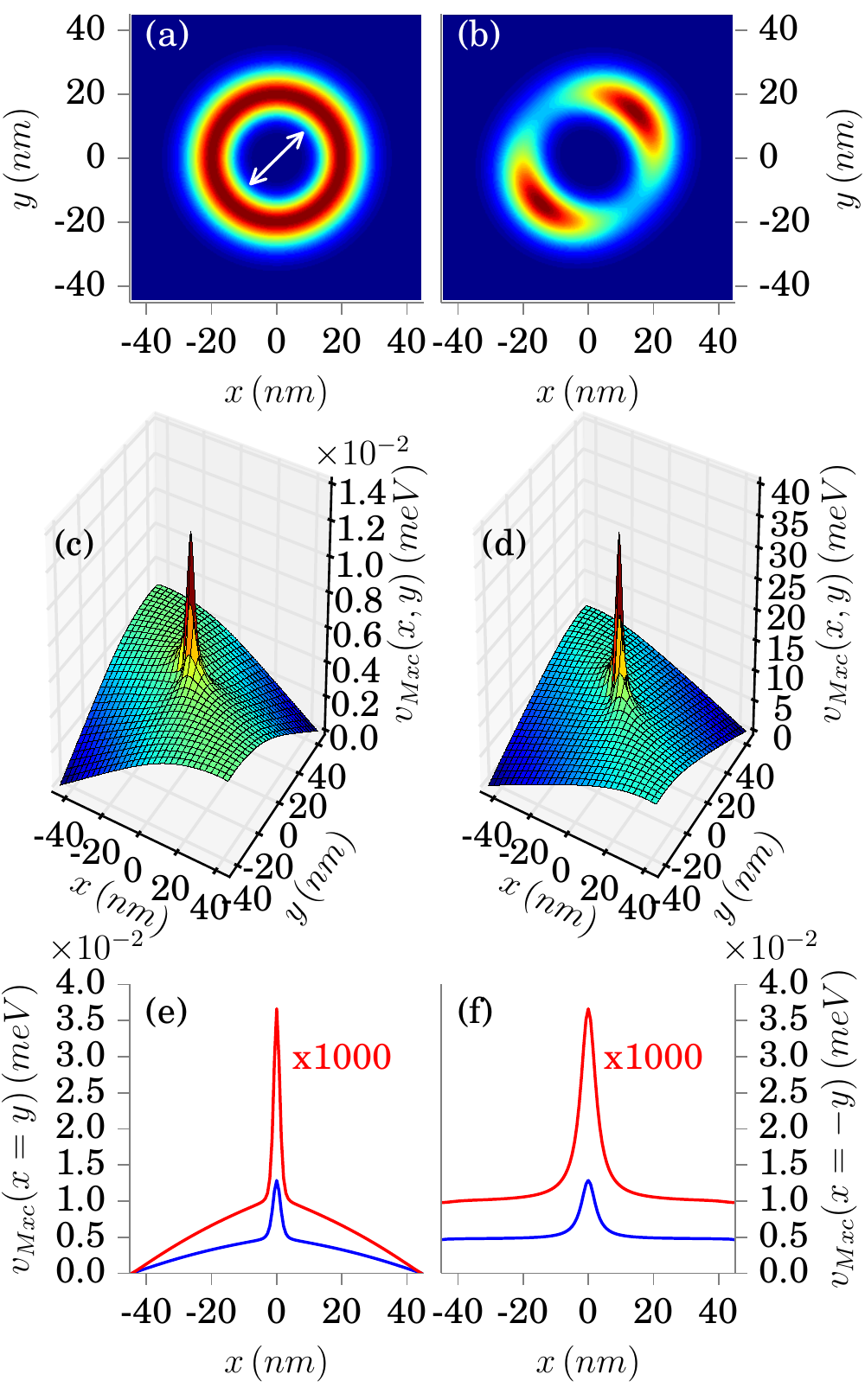}
  \end{center}
  \caption{(Color online) 
Panel (a) shows the ground-state density for {a weak-coupling case with } $\lambda_\alpha = 0.0135$meV$^{1/2}$/nm and {in panel (b) we illustrate a strong-coupling case with} $\lambda_\alpha = 0.54$meV$^{1/2}$/nm. The {corresponding} ground-state Mxc potential for $\lambda_\alpha = 0.0135$meV$^{1/2}$/nm is displayed in (c) and for $\lambda_\alpha = 0.54$meV$^{1/2}$/nm displayed in (d). In (e) and (f) cuts (blue $\lambda_\alpha = 0.0135$meV$^{1/2}$/nm and red $\lambda_\alpha = 0.54$meV$^{1/2}$/nm) through $v_{\mathrm{Mxc}}$ along the diagonal(c) /antidiagonal(d) are shown. The white arrow in (a) indicates the polarization direction of the field mode.}
  \label{fig:static} 
\end{figure}
\begin{figure*}[ht]  
  \begin{center}  
    \includegraphics[width=\textwidth]{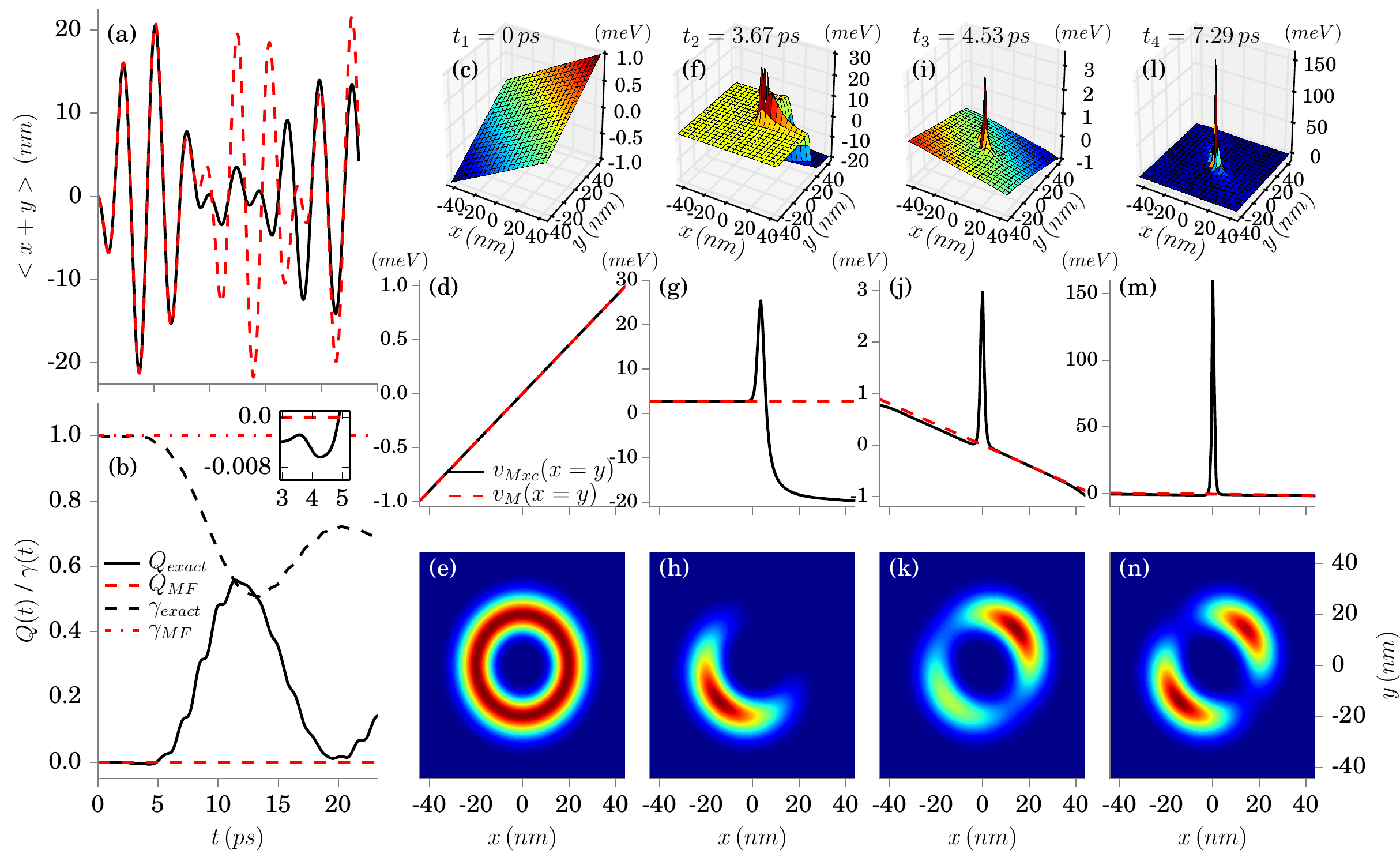}
  \end{center}
 \caption{(Color online) Coherent {state} as initial state {for the photon mode}: In (a) we display the dipole of the exact (black) and mean-field (red) time evolution. In (b) we contrast the exact Mandel parameter $Q(t)$ (see text for definition) (solid black) and $\gamma(t)$ (dotted black) with the corresponding mean-field values (red). In (c), (f), (i) and (l) the corresponding Mxc potentials at different times ($t=0,3.67,4.53,7.29\ ps$), in (d), (g), (j) and (m) the corresponding diagonal cuts of the Mxc potentials, and in (e), (h), (k), and (n) we present the corresponding densities. The inset in (b) shows Q(t) between t=3ps and 5ps. The negative Q(t) in the exact solution indicates nonclassical behavior in the photon mode. Supporting information (SI) Movie 1 shows the full time-evolution from 0ps to 23ps.}
  \label{fig:td-cs}
\end{figure*}
In Fig.~\ref{fig:static}~(a) we show a ground-state density in the weak-coupling case ($\lambda_\alpha = 0.0135$meV$^{1/2}$/nm). Compared to the cavity-free case, we see a slight prolongation of the density along the $x=y$ axis. If we increase the coupling ($\lambda_\alpha = 0.54$meV$^{1/2}$/nm) this feature becomes more dominant and the charge density of the electron becomes even separated (see Fig.~\ref{fig:static}~(b)). If we consider the Mxc potential we {observe} a peak in the middle of the cavity, which models the forces that the photons exert on the electron to elongate its charge distribution (see Fig.~\ref{fig:static}~(c) for weak-coupling and Fig.~\ref{fig:static}~(d) for strong-coupling). In Fig.~\ref{fig:static}~(e) and (f), we show diagonal cuts through $v_\text{Mxc}(\textbf{r})$ for the weak (blue) as well as the strong coupling (red) regime. We see how $v_\text{Mxc}$ pushes the density further apart the stronger the coupling becomes. Such a splitting can not be generated using a static classical field in dipole coupling. Hence the non-dipole contributions of the $v_\text{Mxc}$ mark the non-classical interaction with the photons and are a necessary feature to model the exact forces exerted by the photons on the charged particles. To further substantiate our findings we have determined the photon-number expectation value $\langle \ad_{\alpha} \a_{\alpha}\rangle$ as well as the purity $\gamma = \text{Tr}\left(\rho_{ph}^2 \right)$ of the ground state, where $\rho_{ph}$ is the reduced photon density matrix. Here, a purity value that deviates from $1$ indicates that the state is not factorizable into photon and electron wave functions. {Therefore, the purity is a measure for electron-photon entanglement.} For weak coupling we have $1.18 \cdot 10^{-3}$ photons and $\gamma = 0.999763$ and for strong coupling we find $0.655$ photons in the ground-state with $\gamma = 0.5969$. {This clearly indicates} that the ground state is a hybrid-state of the photons and the electron with stronger entanglement in the strong-coupling regime. A {further} parameter {that} we consider in our analysis is the Mandel $Q$ parameter~\cite{mandel-1979}
\begin{align}
Q = \frac{\langle \ad_{\alpha} \ad_{\alpha} \a_{\alpha} \a_{\alpha} \rangle -  \langle \ad_{\alpha} \a_{\alpha} \rangle^2 }{\langle \ad_{\alpha} \a_{\alpha} \rangle},
\end{align} 
which measures the deviation of the photon statistics from a Poisson distribution {and thus is a measure for the {quantum nature} of the photonic subsystem}. If the field is in a quasi-classical state, i.e., in a coherent state, then $Q=0$. For weak coupling we find $Q= 3.87 \cdot 10^{-4}$, while for strong coupling we have $Q =  0.3361$. This further supports that this model has a highly non-classical ground state of the coupled matter-photon system. 
\\ 
Next, we {turn our attention to the} time-dependent situation. As initial states of the combined matter-photon system (as well as for the corresponding Kohn-Sham system) we {consider two different cases. In both cases we} choose factorizable initial states, which consist of the electronic ground state of the unperturbed quantum ring and the photon field in {(1)} a coherent state with $\langle \ad_{\alpha}\a_{\alpha} \rangle=4$ and in {(2)} a superposition of the vacuum state and the one-photon state with $\langle \ad_{\alpha}\a_{\alpha} \rangle=0.5$. In both examples, we choose the electron-photon coupling strength $\lambda_\alpha = 0.027$meV$^{1/2}$/nm. To numerically propagate the system, we use a Lanczos scheme and propagate the initial state in {(1)} 160000 ({(2) 360000}) timesteps with {$\Delta t = 0.146$fs}. We start with the analysis of example (1). Here, we compare the exact dynamics to {the dynamics} induced by the (self-consistent) classical mean-field approximation. In Fig.~\ref{fig:td-cs}~(a) we display the time evolution of the exact {experimentally accessible} dipole moment (black) and contrast it to the classical approximation (red). Since the field is initially in a coherent state and resembles a classical field, the evolution of the classical approximation is for small times similar to the exact one. Between {$t=0$ps and $t=6$ps}, the classical approximation is the dominant part for most of the time in $v_\text{Mxc}(\textbf{r}t)$ (see Fig.~\ref{fig:td-cs}~(c)). Nevertheless, also in this time interval we find large beyond-dipole corrections to $v_\text{Mxc}(\textbf{r}t)${, which appear} at the turning points of the dipole evolution {and vanish afterwards}. {Precisely} at these times, we observe several peaks and steps in the two-dimensional surface plot {of the effective potential} (see Fig.~\ref{fig:td-cs}~(f) and (i)) that describe the non-classical forces due to the interaction with the photon mode. After { $t=6$ps}, the beyond-dipole correction becomes the dominant part in $v_\text{Mxc}(\textbf{r}t)$ and we find a dominating peak in the two-dimensional surface plot (see Fig.~\ref{fig:td-cs}~(l)). {The peak structure of $v_{Mxc}$ becomes clearly visible in Fig.~\ref{fig:td-cs}~(g), (j), and (m), where we plot the diagonal of $v_{Mxc}$ for the different timesteps.} To further analyze this time-dependent system, we computed the (now time-dependent) Mandel $Q(t)$ parameter and the purity $\gamma(t)$. \\
In Fig.~\ref{fig:td-cs}~(b) we contrast the exact results (black) to those found from the mean-field calculation (red). The purity (dotted black) as well as the Mandel $Q(t)$ parameter (solid black) are in agreement with our previous observations, {namely} that around { $t=6$ps}, where these parameters start to deviate more strongly from the mean-field values, the classical description breaks down. We point out, that in our (decoupled) Kohn-Sham system these parameters are by construction constant and equivalent to the mean-field values, and the Kohn-Sham photon field only changes the number of photons in the coherent state. Hence, the values of these parameters become non-trivial functionals of the initial state as well as $n$ and $q_{\alpha}$. {In particular the assessment of the purity allows us to conclude that the peaks in $v_\text{Mxc}$ are associated with how close to a factorizable (electronic) state the many-body system is. For small times, the system remains close to a factorizable state (purity value close to 1) and we find peaks and steps only at the turning point of the dipole moment, while later {in time} memory effects become dominant and cause permanent peaks and steps.} Finally we note that while we have termed all beyond-dipole contributions to $v_\text{Mxc}(\textbf{r}t)$ as non-classical (since they come solely from the quantum nature of light), the non-classicality of the photon field alone is often associated with a negative $Q(t)$. In Fig.~\ref{fig:td-cs}~(b) we have thus 
provided an inset to highlight that (up to { $t=5$ps, see inset}) such sub-Poissonian statistics, which cannot be described by any probability distribution in phase space, are also present in our prototype system. \\
\begin{figure*}[ht] 
\centerline{\includegraphics[width=\textwidth]{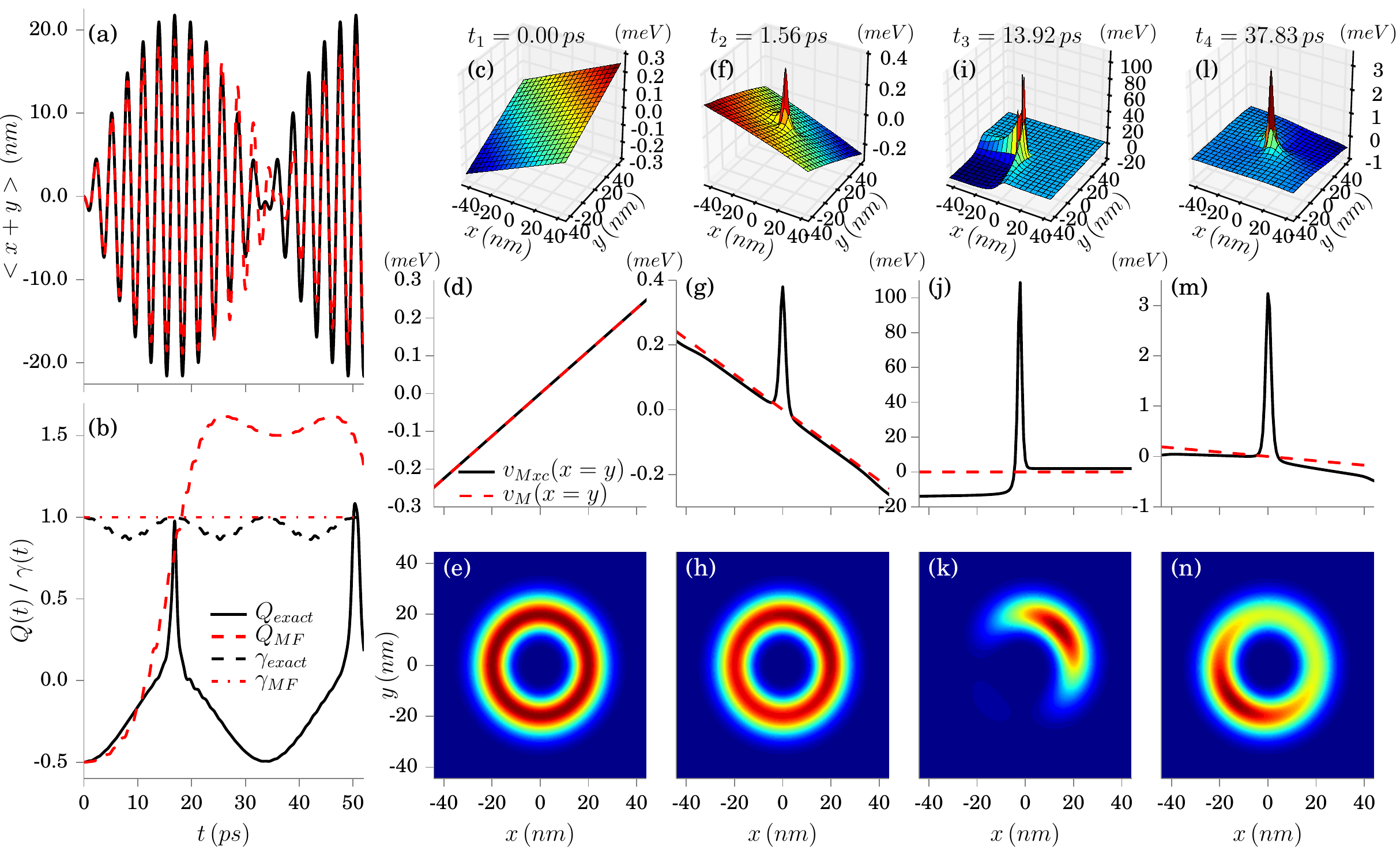}}
 \caption{(Color online) Superposition {of Fock number states as} initial state {for the photon mode}: In (a) we display the dipole of the exact (black) and mean-field (red) time evolution. In (b) we contrast the exact $Q(t)$ (solid black) and $\gamma(t)$ (dotted black) with the corresponding mean-field values (red). In (c), (f), (i) and (l) the corresponding Mxc potentials at different times ($t=0,1.56,13.92,37.83\ ps$), in (d), (g), (j) and (m) the corresponding diagonal cuts of the Mxc potentials, and in (e), (h), (k) and (n) we present the corresponding densities. SI Movie 2 shows the full time-evolution from 0ps to 50ps.}
  \label{fig:td-qs}
\end{figure*}
Next, we analyze in detail our example (2) in Fig.~\ref{fig:td-qs}. We again compare the exact dynamics to {the time-evolution} induced by the (self-consistent) classical mean-field approximation. {For the photon mode we chose in this example as initial state a superposition of the lowest two Fock number states.} Although the field is {in this case} initially in a state {that is not resembling a classical situation as in our first example}, the evolution of the classical approximation is for small times similar to the exact one. Between {$t=0$ps and $t=2$ps}, the classical approximation is the dominant part for most of the time in $v_\text{Mxc}(\textbf{r}t)$ (see Fig.~\ref{fig:td-qs}~(c)). Nevertheless, also in this time interval we find large beyond-dipole corrections to $v_\text{Mxc}(\textbf{r}t)$. {After the first turning points of the dipole evolution, a dominant peak in $v_{Mxc}$ appears (see Fig.~\ref{fig:td-qs}~(f)}. Later, we observe {again} several peaks and steps appearing in the two-dimensional surface plot {of the effective potential} (see Fig.~\ref{fig:td-qs}~(f) and (i)) that describe the non-classical forces due to the interaction with the photon mode. After { $t=2$ps}, the beyond-dipole correction becomes the dominant part in $v_\text{Mxc}(\textbf{r}t)$ and we find a dominating peak in the two-dimensional surface plot. {The peak structure of $v_{Mxc}$ becomes clearly visible in Fig.~\ref{fig:td-qs}~(g), (j),and (m), where we plot the diagonal of $v_{Mxc}$ for the different timesteps.} The purity (dotted black) as well as the Mandel $Q(t)$ parameter (solid black) are in agreement with our previous observations in the coherent state example, that around { $t=2$ps}, where these parameters start to deviate more strongly from the mean-field values, the classical description breaks down. For small times, the system remains close to a factorizable state, while later memory effects become dominant and cause permanent peaks and steps. \\
{Both examples shown in Fig.~\ref{fig:td-cs} and Fig.~\ref{fig:td-qs} nicely illustrate} how a Kohn-Sham approach can exactly describe the different regimes of quantized photon fields that interact with matter. Additionally, we provide in SI Movie 3 a supplementary example of a correlated electron-photon propagation with non-factorizable initial state driven by an external laser pulse.\\
In conclusion, we have presented the {real-space signatures of the} exact effective {potentials} {for} a Kohn-Sham approach to cavity QED. {We have identified step and peak structures which are reminiscent of the steps and peaks in the exact Kohn-Sham potential of traditional DFT, but arise here solely due to the coupling to quantized photon fields.} These effective {potentials} account for the forces that the photons and the electrons exert on each other if we employ an uncoupled Kohn-Sham system to describe the coupled matter-photon system. Provided we have a good approximation to the effective potential \cite{pellegrini-2015}, which includes the here observed peak and step structures due to the non-classical light-matter interaction, the Kohn-Sham approach can be used to perform \textit{ab-initio} calculations of large quantum systems interacting with photons in a high-Q cavity. In this case we have a valuable computational tool for QED chemistry \cite{schwartz-2011,hutchison-2012}, which would open up a new field of research for the electronic-structure community. Besides developing approximations to the Mxc potential, a further important line of research is the extension of the current work to cavities with loss \cite{ruggenthaler-2014}, which is the standard situation in most cavity-QED experiments. {Work along these lines is currently in progress in our group.}   
\\

We acknowledges financial support from the European Research Council Advanced Grant DYNamo (ERC-2010- AdG-267374), Spanish Grant (FIS2013-46159-C3-1-P), Grupos Consolidados UPV/EHU del Gobierno Vasco (IT578-13), COST Actions CM1204 (XLIC), MP1306 (EUSpec) and the Austrian Science {Fund} (FWF P25739-N27).

\bibliographystyle{apsrev4-1}
\bibliography{fixed_point_letter}

\begin{thebibliography}{31}%
\makeatletter
\providecommand \@ifxundefined [1]{%
 \@ifx{#1\undefined}
}%
\providecommand \@ifnum [1]{%
 \ifnum #1\expandafter \@firstoftwo
 \else \expandafter \@secondoftwo
 \fi
}%
\providecommand \@ifx [1]{%
 \ifx #1\expandafter \@firstoftwo
 \else \expandafter \@secondoftwo
 \fi
}%
\providecommand \natexlab [1]{#1}%
\providecommand \enquote  [1]{``#1''}%
\providecommand \bibnamefont  [1]{#1}%
\providecommand \bibfnamefont [1]{#1}%
\providecommand \citenamefont [1]{#1}%
\providecommand \href@noop [0]{\@secondoftwo}%
\providecommand \href [0]{\begingroup \@sanitize@url \@href}%
\providecommand \@href[1]{\@@startlink{#1}\@@href}%
\providecommand \@@href[1]{\endgroup#1\@@endlink}%
\providecommand \@sanitize@url [0]{\catcode `\\12\catcode `\$12\catcode
  `\&12\catcode `\#12\catcode `\^12\catcode `\_12\catcode `\%12\relax}%
\providecommand \@@startlink[1]{}%
\providecommand \@@endlink[0]{}%
\providecommand \url  [0]{\begingroup\@sanitize@url \@url }%
\providecommand \@url [1]{\endgroup\@href {#1}{\urlprefix }}%
\providecommand \urlprefix  [0]{URL }%
\providecommand \Eprint [0]{\href }%
\providecommand \doibase [0]{http://dx.doi.org/}%
\providecommand \selectlanguage [0]{\@gobble}%
\providecommand \bibinfo  [0]{\@secondoftwo}%
\providecommand \bibfield  [0]{\@secondoftwo}%
\providecommand \translation [1]{[#1]}%
\providecommand \BibitemOpen [0]{}%
\providecommand \bibitemStop [0]{}%
\providecommand \bibitemNoStop [0]{.\EOS\space}%
\providecommand \EOS [0]{\spacefactor3000\relax}%
\providecommand \BibitemShut  [1]{\csname bibitem#1\endcsname}%
\let\auto@bib@innerbib\@empty
%</preamble>
\bibitem [{\citenamefont {Kippenberg}\ and\ \citenamefont
  {Vahala}(2008)}]{kippenberg-2008}%
  \BibitemOpen
  \bibfield  {author} {\bibinfo {author} {\bibfnamefont {T.~J.}\ \bibnamefont
  {Kippenberg}}\ and\ \bibinfo {author} {\bibfnamefont {K.~J.}\ \bibnamefont
  {Vahala}},\ }\href {\doibase 10.1126/science.1156032} {\bibfield  {journal}
  {\bibinfo  {journal} {Science}\ }\textbf {\bibinfo {volume} {321}},\ \bibinfo
  {pages} {1172} (\bibinfo {year} {2008})}\BibitemShut {NoStop}%
\bibitem [{\citenamefont {Schoelkopf}\ and\ \citenamefont
  {Girvin}(2008)}]{schoelkopf-2008}%
  \BibitemOpen
  \bibfield  {author} {\bibinfo {author} {\bibfnamefont {R.~J.}\ \bibnamefont
  {Schoelkopf}}\ and\ \bibinfo {author} {\bibfnamefont {S.~M.}\ \bibnamefont
  {Girvin}},\ }\href {\doibase 10.1038/451664a} {\bibfield  {journal} {\bibinfo
   {journal} {Nature}\ }\textbf {\bibinfo {volume} {451}},\ \bibinfo {pages}
  {664} (\bibinfo {year} {2008})}\BibitemShut {NoStop}%
\bibitem [{\citenamefont {Giovannini}\ \emph {et~al.}(2015)\citenamefont
  {Giovannini}, \citenamefont {Romero}, \citenamefont {Potoček}, \citenamefont
  {Ferenczi}, \citenamefont {Speirits}, \citenamefont {Barnett}, \citenamefont
  {Faccio},\ and\ \citenamefont {Padgett}}]{giovannini-2015}%
  \BibitemOpen
  \bibfield  {author} {\bibinfo {author} {\bibfnamefont {D.}~\bibnamefont
  {Giovannini}}, \bibinfo {author} {\bibfnamefont {J.}~\bibnamefont {Romero}},
  \bibinfo {author} {\bibfnamefont {V.}~\bibnamefont {Potoček}}, \bibinfo
  {author} {\bibfnamefont {G.}~\bibnamefont {Ferenczi}}, \bibinfo {author}
  {\bibfnamefont {F.}~\bibnamefont {Speirits}}, \bibinfo {author}
  {\bibfnamefont {S.~M.}\ \bibnamefont {Barnett}}, \bibinfo {author}
  {\bibfnamefont {D.}~\bibnamefont {Faccio}}, \ and\ \bibinfo {author}
  {\bibfnamefont {M.~J.}\ \bibnamefont {Padgett}},\ }\href {\doibase
  10.1126/science.aaa3035} {\bibfield  {journal} {\bibinfo  {journal}
  {Science}\ }\textbf {\bibinfo {volume} {347}},\ \bibinfo {pages} {857}
  (\bibinfo {year} {2015})}\BibitemShut {NoStop}%
\bibitem [{\citenamefont {Casabone}\ \emph {et~al.}(2015)\citenamefont
  {Casabone}, \citenamefont {Friebe}, \citenamefont {Brandst\"atter},
  \citenamefont {Sch\"uppert}, \citenamefont {Blatt},\ and\ \citenamefont
  {Northup}}]{casabone-2015}%
  \BibitemOpen
  \bibfield  {author} {\bibinfo {author} {\bibfnamefont {B.}~\bibnamefont
  {Casabone}}, \bibinfo {author} {\bibfnamefont {K.}~\bibnamefont {Friebe}},
  \bibinfo {author} {\bibfnamefont {B.}~\bibnamefont {Brandst\"atter}},
  \bibinfo {author} {\bibfnamefont {K.}~\bibnamefont {Sch\"uppert}}, \bibinfo
  {author} {\bibfnamefont {R.}~\bibnamefont {Blatt}}, \ and\ \bibinfo {author}
  {\bibfnamefont {T.~E.}\ \bibnamefont {Northup}},\ }\href {\doibase
  10.1103/PhysRevLett.114.023602} {\bibfield  {journal} {\bibinfo  {journal}
  {Phys. Rev. Lett.}\ }\textbf {\bibinfo {volume} {114}},\ \bibinfo {pages}
  {023602} (\bibinfo {year} {2015})}\BibitemShut {NoStop}%
\bibitem [{\citenamefont {Schwartz}\ \emph {et~al.}(2011)\citenamefont
  {Schwartz}, \citenamefont {Hutchison}, \citenamefont {Genet},\ and\
  \citenamefont {Ebbesen}}]{schwartz-2011}%
  \BibitemOpen
  \bibfield  {author} {\bibinfo {author} {\bibfnamefont {T.}~\bibnamefont
  {Schwartz}}, \bibinfo {author} {\bibfnamefont {J.~A.}\ \bibnamefont
  {Hutchison}}, \bibinfo {author} {\bibfnamefont {C.}~\bibnamefont {Genet}}, \
  and\ \bibinfo {author} {\bibfnamefont {T.~W.}\ \bibnamefont {Ebbesen}},\
  }\href {\doibase 10.1103/PhysRevLett.106.196405} {\bibfield  {journal}
  {\bibinfo  {journal} {Phys. Rev. Lett.}\ }\textbf {\bibinfo {volume} {106}},\
  \bibinfo {pages} {196405} (\bibinfo {year} {2011})}\BibitemShut {NoStop}%
\bibitem [{\citenamefont {Hutchison}\ \emph {et~al.}(2012)\citenamefont
  {Hutchison}, \citenamefont {Schwartz}, \citenamefont {Genet}, \citenamefont
  {Devaux},\ and\ \citenamefont {Ebbesen}}]{hutchison-2012}%
  \BibitemOpen
  \bibfield  {author} {\bibinfo {author} {\bibfnamefont {J.~A.}\ \bibnamefont
  {Hutchison}}, \bibinfo {author} {\bibfnamefont {T.}~\bibnamefont {Schwartz}},
  \bibinfo {author} {\bibfnamefont {C.}~\bibnamefont {Genet}}, \bibinfo
  {author} {\bibfnamefont {E.}~\bibnamefont {Devaux}}, \ and\ \bibinfo {author}
  {\bibfnamefont {T.~W.}\ \bibnamefont {Ebbesen}},\ }\href {\doibase
  10.1002/anie.201107033} {\bibfield  {journal} {\bibinfo  {journal}
  {Angewandte Chemie International Edition}\ }\textbf {\bibinfo {volume}
  {51}},\ \bibinfo {pages} {1592} (\bibinfo {year} {2012})}\BibitemShut
  {NoStop}%
\bibitem [{\citenamefont {Fetter}\ and\ \citenamefont
  {Walecka}(2003)}]{fetter-2003}%
  \BibitemOpen
  \bibfield  {author} {\bibinfo {author} {\bibfnamefont {A.~L.}\ \bibnamefont
  {Fetter}}\ and\ \bibinfo {author} {\bibfnamefont {J.~D.}\ \bibnamefont
  {Walecka}},\ }\href@noop {} {\emph {\bibinfo {title} {Quantum theory of
  many-particle systems}}}\ (\bibinfo  {publisher} {Courier Dover
  Publications},\ \bibinfo {year} {2003})\BibitemShut {NoStop}%
\bibitem [{\citenamefont {Marques}\ \emph {et~al.}(2012)\citenamefont
  {Marques}, \citenamefont {Maitra}, \citenamefont {Nogueira}, \citenamefont
  {Gross},\ and\ \citenamefont {Rubio}}]{marques-2012}%
  \BibitemOpen
  \bibfield  {author} {\bibinfo {author} {\bibfnamefont {M.~A.}\ \bibnamefont
  {Marques}}, \bibinfo {author} {\bibfnamefont {N.~T.}\ \bibnamefont {Maitra}},
  \bibinfo {author} {\bibfnamefont {F.~M.}\ \bibnamefont {Nogueira}}, \bibinfo
  {author} {\bibfnamefont {E.~K.}\ \bibnamefont {Gross}}, \ and\ \bibinfo
  {author} {\bibfnamefont {A.}~\bibnamefont {Rubio}},\ }\href@noop {} {\emph
  {\bibinfo {title} {Fundamentals of time-dependent density functional
  theory}}},\ Vol.\ \bibinfo {volume} {837}\ (\bibinfo  {publisher}
  {Springer},\ \bibinfo {year} {2012})\BibitemShut {NoStop}%
\bibitem [{\citenamefont {Ullrich}(2012)}]{ullrich-2012}%
  \BibitemOpen
  \bibfield  {author} {\bibinfo {author} {\bibfnamefont {C.~A.}\ \bibnamefont
  {Ullrich}},\ }\href@noop {} {\emph {\bibinfo {title} {Time-dependent
  density-functional theory: concepts and applications}}}\ (\bibinfo
  {publisher} {Oxford University Press},\ \bibinfo {year} {2012})\BibitemShut
  {NoStop}%
\bibitem [{\citenamefont {Stefanucci}\ and\ \citenamefont {van
  Leeuwen}(2013)}]{stefanucci-2013}%
  \BibitemOpen
  \bibfield  {author} {\bibinfo {author} {\bibfnamefont {G.}~\bibnamefont
  {Stefanucci}}\ and\ \bibinfo {author} {\bibfnamefont {R.}~\bibnamefont {van
  Leeuwen}},\ }\href@noop {} {\emph {\bibinfo {title} {Nonequilibrium Many-Body
  Theory of Quantum Systems: A Modern Introduction}}}\ (\bibinfo  {publisher}
  {Cambridge University Press},\ \bibinfo {year} {2013})\BibitemShut {NoStop}%
\bibitem [{\citenamefont {Bartlett}\ and\ \citenamefont
  {Musia\l{}}(2007)}]{Bartlett-2007}%
  \BibitemOpen
  \bibfield  {author} {\bibinfo {author} {\bibfnamefont {R.~J.}\ \bibnamefont
  {Bartlett}}\ and\ \bibinfo {author} {\bibfnamefont {M.}~\bibnamefont
  {Musia\l{}}},\ }\href {\doibase 10.1103/RevModPhys.79.291} {\bibfield
  {journal} {\bibinfo  {journal} {Rev. Mod. Phys.}\ }\textbf {\bibinfo {volume}
  {79}},\ \bibinfo {pages} {291} (\bibinfo {year} {2007})}\BibitemShut
  {NoStop}%
\bibitem [{\citenamefont {Thomas}\ \emph {et~al.}(2015)\citenamefont {Thomas},
  \citenamefont {Booth},\ and\ \citenamefont {Alavi}}]{Thomas-2015}%
  \BibitemOpen
  \bibfield  {author} {\bibinfo {author} {\bibfnamefont {R.~E.}\ \bibnamefont
  {Thomas}}, \bibinfo {author} {\bibfnamefont {G.~H.}\ \bibnamefont {Booth}}, \
  and\ \bibinfo {author} {\bibfnamefont {A.}~\bibnamefont {Alavi}},\ }\href
  {\doibase 10.1103/PhysRevLett.114.033001} {\bibfield  {journal} {\bibinfo
  {journal} {Phys. Rev. Lett.}\ }\textbf {\bibinfo {volume} {114}},\ \bibinfo
  {pages} {033001} (\bibinfo {year} {2015})}\BibitemShut {NoStop}%
\bibitem [{\citenamefont {Ruggenthaler}\ \emph {et~al.}(2011)\citenamefont
  {Ruggenthaler}, \citenamefont {Mackenroth},\ and\ \citenamefont
  {Bauer}}]{ruggenthaler-2011b}%
  \BibitemOpen
  \bibfield  {author} {\bibinfo {author} {\bibfnamefont {M.}~\bibnamefont
  {Ruggenthaler}}, \bibinfo {author} {\bibfnamefont {F.}~\bibnamefont
  {Mackenroth}}, \ and\ \bibinfo {author} {\bibfnamefont {D.}~\bibnamefont
  {Bauer}},\ }\href {\doibase 10.1103/PhysRevA.84.042107} {\bibfield  {journal}
  {\bibinfo  {journal} {Phys. Rev. A}\ }\textbf {\bibinfo {volume} {84}},\
  \bibinfo {pages} {042107} (\bibinfo {year} {2011})}\BibitemShut {NoStop}%
\bibitem [{\citenamefont {Tokatly}(2013)}]{tokatly-2013}%
  \BibitemOpen
  \bibfield  {author} {\bibinfo {author} {\bibfnamefont {I.~V.}\ \bibnamefont
  {Tokatly}},\ }\href {\doibase 10.1103/PhysRevLett.110.233001} {\bibfield
  {journal} {\bibinfo  {journal} {Phys. Rev. Lett.}\ }\textbf {\bibinfo
  {volume} {110}},\ \bibinfo {pages} {233001} (\bibinfo {year}
  {2013})}\BibitemShut {NoStop}%
\bibitem [{\citenamefont {Ruggenthaler}\ \emph {et~al.}(2014)\citenamefont
  {Ruggenthaler}, \citenamefont {Flick}, \citenamefont {Pellegrini},
  \citenamefont {Appel}, \citenamefont {Tokatly},\ and\ \citenamefont
  {Rubio}}]{ruggenthaler-2014}%
  \BibitemOpen
  \bibfield  {author} {\bibinfo {author} {\bibfnamefont {M.}~\bibnamefont
  {Ruggenthaler}}, \bibinfo {author} {\bibfnamefont {J.}~\bibnamefont {Flick}},
  \bibinfo {author} {\bibfnamefont {C.}~\bibnamefont {Pellegrini}}, \bibinfo
  {author} {\bibfnamefont {H.}~\bibnamefont {Appel}}, \bibinfo {author}
  {\bibfnamefont {I.~V.}\ \bibnamefont {Tokatly}}, \ and\ \bibinfo {author}
  {\bibfnamefont {A.}~\bibnamefont {Rubio}},\ }\href {\doibase
  10.1103/PhysRevA.90.012508} {\bibfield  {journal} {\bibinfo  {journal} {Phys.
  Rev. A}\ }\textbf {\bibinfo {volume} {90}},\ \bibinfo {pages} {012508}
  (\bibinfo {year} {2014})}\BibitemShut {NoStop}%
\bibitem [{\citenamefont {Marques}\ \emph {et~al.}(2003)\citenamefont
  {Marques}, \citenamefont {Castro}, \citenamefont {Bertsch},\ and\
  \citenamefont {Rubio}}]{marques-2003}%
  \BibitemOpen
  \bibfield  {author} {\bibinfo {author} {\bibfnamefont {M.~A.}\ \bibnamefont
  {Marques}}, \bibinfo {author} {\bibfnamefont {A.}~\bibnamefont {Castro}},
  \bibinfo {author} {\bibfnamefont {G.~F.}\ \bibnamefont {Bertsch}}, \ and\
  \bibinfo {author} {\bibfnamefont {A.}~\bibnamefont {Rubio}},\ }\href
  {\doibase http://dx.doi.org/10.1016/S0010-4655(02)00686-0} {\bibfield
  {journal} {\bibinfo  {journal} {Computer Physics Communications}\ }\textbf
  {\bibinfo {volume} {151}},\ \bibinfo {pages} {60 } (\bibinfo {year}
  {2003})}\BibitemShut {NoStop}%
\bibitem [{\citenamefont {Castro}\ \emph {et~al.}(2006)\citenamefont {Castro},
  \citenamefont {Appel}, \citenamefont {Oliveira}, \citenamefont {Rozzi},
  \citenamefont {Andrade}, \citenamefont {Lorenzen}, \citenamefont {Marques},
  \citenamefont {Gross},\ and\ \citenamefont {Rubio}}]{castro-2006}%
  \BibitemOpen
  \bibfield  {author} {\bibinfo {author} {\bibfnamefont {A.}~\bibnamefont
  {Castro}}, \bibinfo {author} {\bibfnamefont {H.}~\bibnamefont {Appel}},
  \bibinfo {author} {\bibfnamefont {M.}~\bibnamefont {Oliveira}}, \bibinfo
  {author} {\bibfnamefont {C.~A.}\ \bibnamefont {Rozzi}}, \bibinfo {author}
  {\bibfnamefont {X.}~\bibnamefont {Andrade}}, \bibinfo {author} {\bibfnamefont
  {F.}~\bibnamefont {Lorenzen}}, \bibinfo {author} {\bibfnamefont {M.~A.~L.}\
  \bibnamefont {Marques}}, \bibinfo {author} {\bibfnamefont {E.~K.~U.}\
  \bibnamefont {Gross}}, \ and\ \bibinfo {author} {\bibfnamefont
  {A.}~\bibnamefont {Rubio}},\ }\href {\doibase 10.1002/pssb.200642067}
  {\bibfield  {journal} {\bibinfo  {journal} {physica status solidi (b)}\
  }\textbf {\bibinfo {volume} {243}},\ \bibinfo {pages} {2465} (\bibinfo {year}
  {2006})}\BibitemShut {NoStop}%
\bibitem [{\citenamefont {Andrade}\ \emph {et~al.}(2015)\citenamefont
  {Andrade}, \citenamefont {Strubbe}, \citenamefont {De~Giovannini},
  \citenamefont {Larsen}, \citenamefont {Oliveira}, \citenamefont
  {Alberdi-Rodriguez}, \citenamefont {Varas}, \citenamefont {Theophilou},
  \citenamefont {Helbig}, \citenamefont {Verstraete}, \citenamefont {Stella},
  \citenamefont {Nogueira}, \citenamefont {Aspuru-Guzik}, \citenamefont
  {Castro}, \citenamefont {Marques},\ and\ \citenamefont
  {Rubio}}]{andrade-2014}%
  \BibitemOpen
  \bibfield  {author} {\bibinfo {author} {\bibfnamefont {X.}~\bibnamefont
  {Andrade}}, \bibinfo {author} {\bibfnamefont {D.}~\bibnamefont {Strubbe}},
  \bibinfo {author} {\bibfnamefont {U.}~\bibnamefont {De~Giovannini}}, \bibinfo
  {author} {\bibfnamefont {A.~H.}\ \bibnamefont {Larsen}}, \bibinfo {author}
  {\bibfnamefont {M.~J.~T.}\ \bibnamefont {Oliveira}}, \bibinfo {author}
  {\bibfnamefont {J.}~\bibnamefont {Alberdi-Rodriguez}}, \bibinfo {author}
  {\bibfnamefont {A.}~\bibnamefont {Varas}}, \bibinfo {author} {\bibfnamefont
  {I.}~\bibnamefont {Theophilou}}, \bibinfo {author} {\bibfnamefont
  {N.}~\bibnamefont {Helbig}}, \bibinfo {author} {\bibfnamefont {M.~J.}\
  \bibnamefont {Verstraete}}, \bibinfo {author} {\bibfnamefont
  {L.}~\bibnamefont {Stella}}, \bibinfo {author} {\bibfnamefont
  {F.}~\bibnamefont {Nogueira}}, \bibinfo {author} {\bibfnamefont
  {A.}~\bibnamefont {Aspuru-Guzik}}, \bibinfo {author} {\bibfnamefont
  {A.}~\bibnamefont {Castro}}, \bibinfo {author} {\bibfnamefont {M.~A.~L.}\
  \bibnamefont {Marques}}, \ and\ \bibinfo {author} {\bibfnamefont
  {A.}~\bibnamefont {Rubio}},\ }\href {\doibase 10.1039/C5CP00351B} {\bibfield
  {journal} {\bibinfo  {journal} {Phys. Chem. Chem. Phys.}\ } (\bibinfo {year}
  {2015}),\ 10.1039/C5CP00351B}\BibitemShut {NoStop}%
\bibitem [{\citenamefont {Gross}\ and\ \citenamefont
  {Dreizler}(1995)}]{gross-1995}%
  \BibitemOpen
  \bibfield  {author} {\bibinfo {author} {\bibfnamefont {E.~K.}\ \bibnamefont
  {Gross}}\ and\ \bibinfo {author} {\bibfnamefont {R.~M.}\ \bibnamefont
  {Dreizler}},\ }\href@noop {} {\emph {\bibinfo {title} {Density functional
  theory}}}\ (\bibinfo  {publisher} {Springer},\ \bibinfo {year}
  {1995})\BibitemShut {NoStop}%
\bibitem [{\citenamefont {Pellegrini}\ \emph {et~al.}(2015)\citenamefont
  {Pellegrini}, \citenamefont {Flick}, \citenamefont {Tokatly}, \citenamefont
  {Appel},\ and\ \citenamefont {Rubio}}]{pellegrini-2015}%
  \BibitemOpen
  \bibfield  {author} {\bibinfo {author} {\bibfnamefont {C.}~\bibnamefont
  {Pellegrini}}, \bibinfo {author} {\bibfnamefont {J.}~\bibnamefont {Flick}},
  \bibinfo {author} {\bibfnamefont {I.~V.}\ \bibnamefont {Tokatly}}, \bibinfo
  {author} {\bibfnamefont {H.}~\bibnamefont {Appel}}, \ and\ \bibinfo {author}
  {\bibfnamefont {A.}~\bibnamefont {Rubio}},\ }\href {\doibase
  10.1103/PhysRevLett.115.093001} {\bibfield  {journal} {\bibinfo  {journal}
  {Phys. Rev. Lett.}\ }\textbf {\bibinfo {volume} {115}},\ \bibinfo {pages}
  {093001} (\bibinfo {year} {2015})}\BibitemShut {NoStop}%
\bibitem [{\citenamefont {Ruggenthaler}\ and\ \citenamefont {van
  Leeuwen}(2011)}]{ruggenthaler-2011a}%
  \BibitemOpen
  \bibfield  {author} {\bibinfo {author} {\bibfnamefont {M.}~\bibnamefont
  {Ruggenthaler}}\ and\ \bibinfo {author} {\bibfnamefont {R.}~\bibnamefont {van
  Leeuwen}},\ }\href {http://stacks.iop.org/0295-5075/95/i=1/a=13001}
  {\bibfield  {journal} {\bibinfo  {journal} {EPL (Europhysics Letters)}\
  }\textbf {\bibinfo {volume} {95}},\ \bibinfo {pages} {13001} (\bibinfo {year}
  {2011})}\BibitemShut {NoStop}%
\bibitem [{\citenamefont {Ruggenthaler}\ \emph {et~al.}(2012)\citenamefont
  {Ruggenthaler}, \citenamefont {Giesbertz}, \citenamefont {Penz},\ and\
  \citenamefont {van Leeuwen}}]{ruggenthaler-2012}%
  \BibitemOpen
  \bibfield  {author} {\bibinfo {author} {\bibfnamefont {M.}~\bibnamefont
  {Ruggenthaler}}, \bibinfo {author} {\bibfnamefont {K.~J.~H.}\ \bibnamefont
  {Giesbertz}}, \bibinfo {author} {\bibfnamefont {M.}~\bibnamefont {Penz}}, \
  and\ \bibinfo {author} {\bibfnamefont {R.}~\bibnamefont {van Leeuwen}},\
  }\href {\doibase 10.1103/PhysRevA.85.052504} {\bibfield  {journal} {\bibinfo
  {journal} {Phys. Rev. A}\ }\textbf {\bibinfo {volume} {85}},\ \bibinfo
  {pages} {052504} (\bibinfo {year} {2012})}\BibitemShut {NoStop}%
\bibitem [{\citenamefont {Nielsen}\ \emph {et~al.}(2013)\citenamefont
  {Nielsen}, \citenamefont {Ruggenthaler},\ and\ \citenamefont {van
  Leeuwen}}]{nielsen-2013}%
  \BibitemOpen
  \bibfield  {author} {\bibinfo {author} {\bibfnamefont {S.~E.~B.}\
  \bibnamefont {Nielsen}}, \bibinfo {author} {\bibfnamefont {M.}~\bibnamefont
  {Ruggenthaler}}, \ and\ \bibinfo {author} {\bibfnamefont {R.}~\bibnamefont
  {van Leeuwen}},\ }\href {http://stacks.iop.org/0295-5075/101/i=3/a=33001}
  {\bibfield  {journal} {\bibinfo  {journal} {EPL (Europhysics Letters)}\
  }\textbf {\bibinfo {volume} {101}},\ \bibinfo {pages} {33001} (\bibinfo
  {year} {2013})}\BibitemShut {NoStop}%
\bibitem [{\citenamefont {{Nielsen}}\ \emph {et~al.}(2014)\citenamefont
  {{Nielsen}}, \citenamefont {{Ruggenthaler}},\ and\ \citenamefont {{van
  Leeuwen}}}]{nielsen-2015}%
  \BibitemOpen
  \bibfield  {author} {\bibinfo {author} {\bibfnamefont {S.~E.~B.}\
  \bibnamefont {{Nielsen}}}, \bibinfo {author} {\bibfnamefont {M.}~\bibnamefont
  {{Ruggenthaler}}}, \ and\ \bibinfo {author} {\bibfnamefont {R.}~\bibnamefont
  {{van Leeuwen}}},\ }\href@noop {} {\bibfield  {journal} {\bibinfo  {journal}
  {ArXiv e-prints}\ } (\bibinfo {year} {2014})},\ \Eprint
  {http://arxiv.org/abs/1412.3794} {arXiv:1412.3794 [quant-ph]} \BibitemShut
  {NoStop}%
\bibitem [{\citenamefont {Elliott}\ \emph {et~al.}(2012)\citenamefont
  {Elliott}, \citenamefont {Fuks}, \citenamefont {Rubio},\ and\ \citenamefont
  {Maitra}}]{elliott-2012}%
  \BibitemOpen
  \bibfield  {author} {\bibinfo {author} {\bibfnamefont {P.}~\bibnamefont
  {Elliott}}, \bibinfo {author} {\bibfnamefont {J.~I.}\ \bibnamefont {Fuks}},
  \bibinfo {author} {\bibfnamefont {A.}~\bibnamefont {Rubio}}, \ and\ \bibinfo
  {author} {\bibfnamefont {N.~T.}\ \bibnamefont {Maitra}},\ }\href {\doibase
  10.1103/PhysRevLett.109.266404} {\bibfield  {journal} {\bibinfo  {journal}
  {Phys. Rev. Lett.}\ }\textbf {\bibinfo {volume} {109}},\ \bibinfo {pages}
  {266404} (\bibinfo {year} {2012})}\BibitemShut {NoStop}%
\bibitem [{\citenamefont {Fuks}\ \emph {et~al.}(2013)\citenamefont {Fuks},
  \citenamefont {Elliott}, \citenamefont {Rubio},\ and\ \citenamefont
  {Maitra}}]{fuks-2013}%
  \BibitemOpen
  \bibfield  {author} {\bibinfo {author} {\bibfnamefont {J.~I.}\ \bibnamefont
  {Fuks}}, \bibinfo {author} {\bibfnamefont {P.}~\bibnamefont {Elliott}},
  \bibinfo {author} {\bibfnamefont {A.}~\bibnamefont {Rubio}}, \ and\ \bibinfo
  {author} {\bibfnamefont {N.~T.}\ \bibnamefont {Maitra}},\ }\href {\doibase
  10.1021/jz302099f} {\bibfield  {journal} {\bibinfo  {journal} {The Journal of
  Physical Chemistry Letters}\ }\textbf {\bibinfo {volume} {4}},\ \bibinfo
  {pages} {735} (\bibinfo {year} {2013})}\BibitemShut {NoStop}%
\bibitem [{\citenamefont {R\"as\"anen}\ \emph {et~al.}(2007)\citenamefont
  {R\"as\"anen}, \citenamefont {Castro}, \citenamefont {Werschnik},
  \citenamefont {Rubio},\ and\ \citenamefont {Gross}}]{rasanen-2007}%
  \BibitemOpen
  \bibfield  {author} {\bibinfo {author} {\bibfnamefont {E.}~\bibnamefont
  {R\"as\"anen}}, \bibinfo {author} {\bibfnamefont {A.}~\bibnamefont {Castro}},
  \bibinfo {author} {\bibfnamefont {J.}~\bibnamefont {Werschnik}}, \bibinfo
  {author} {\bibfnamefont {A.}~\bibnamefont {Rubio}}, \ and\ \bibinfo {author}
  {\bibfnamefont {E.~K.~U.}\ \bibnamefont {Gross}},\ }\href {\doibase
  10.1103/PhysRevLett.98.157404} {\bibfield  {journal} {\bibinfo  {journal}
  {Phys. Rev. Lett.}\ }\textbf {\bibinfo {volume} {98}},\ \bibinfo {pages}
  {157404} (\bibinfo {year} {2007})}\BibitemShut {NoStop}%
\bibitem [{\citenamefont {Garza}\ and\ \citenamefont
  {Scuseria}(2012)}]{garza-2012}%
  \BibitemOpen
  \bibfield  {author} {\bibinfo {author} {\bibfnamefont {A.~J.}\ \bibnamefont
  {Garza}}\ and\ \bibinfo {author} {\bibfnamefont {G.~E.}\ \bibnamefont
  {Scuseria}},\ }\href {\doibase http://dx.doi.org/10.1063/1.4740249}
  {\bibfield  {journal} {\bibinfo  {journal} {The Journal of Chemical Physics}\
  }\textbf {\bibinfo {volume} {137}},\ \bibinfo {pages} {054110} (\bibinfo
  {year} {2012})}\BibitemShut {NoStop}%
\bibitem [{\citenamefont {Helbig}\ \emph {et~al.}(2009)\citenamefont {Helbig},
  \citenamefont {Tokatly},\ and\ \citenamefont {Rubio}}]{helbig-2009}%
  \BibitemOpen
  \bibfield  {author} {\bibinfo {author} {\bibfnamefont {N.}~\bibnamefont
  {Helbig}}, \bibinfo {author} {\bibfnamefont {I.~V.}\ \bibnamefont {Tokatly}},
  \ and\ \bibinfo {author} {\bibfnamefont {A.}~\bibnamefont {Rubio}},\ }\href
  {\doibase http://dx.doi.org/10.1063/1.3271392} {\bibfield  {journal}
  {\bibinfo  {journal} {The Journal of Chemical Physics}\ }\textbf {\bibinfo
  {volume} {131}},\ \bibinfo {pages} {224105} (\bibinfo {year}
  {2009})}\BibitemShut {NoStop}%
\bibitem [{\citenamefont {Flick}\ \emph {et~al.}(2014)\citenamefont {Flick},
  \citenamefont {Appel},\ and\ \citenamefont {Rubio}}]{flick-2014}%
  \BibitemOpen
  \bibfield  {author} {\bibinfo {author} {\bibfnamefont {J.}~\bibnamefont
  {Flick}}, \bibinfo {author} {\bibfnamefont {H.}~\bibnamefont {Appel}}, \ and\
  \bibinfo {author} {\bibfnamefont {A.}~\bibnamefont {Rubio}},\ }\href
  {\doibase 10.1021/ct4010933} {\bibfield  {journal} {\bibinfo  {journal}
  {Journal of Chemical Theory and Computation}\ }\textbf {\bibinfo {volume}
  {10}},\ \bibinfo {pages} {1665} (\bibinfo {year} {2014})}\BibitemShut
  {NoStop}%
\bibitem [{\citenamefont {Mandel}(1979)}]{mandel-1979}%
  \BibitemOpen
  \bibfield  {author} {\bibinfo {author} {\bibfnamefont {L.}~\bibnamefont
  {Mandel}},\ }\href {\doibase 10.1364/OL.4.000205} {\bibfield  {journal}
  {\bibinfo  {journal} {Opt. Lett.}\ }\textbf {\bibinfo {volume} {4}},\
  \bibinfo {pages} {205} (\bibinfo {year} {1979})}\BibitemShut {NoStop}%
\end{thebibliography}%

\end{document}